\documentclass[a4paper,12pt]{article} 
\usepackage[utf8]{inputenc} 
\usepackage[english]{babel} 
\usepackage[hyperindex,breaklinks]{hyperref} 
\usepackage{graphicx}
 
\title{The women day storm}
\author{Aleksei~Parnowski\thanks{Corresponding author; e-mail: parnowski@ikd.kiev.ua / 
fax: +380(44)5264124}, Anna~Polonska and Oleg~Semeniv \\ Space Research Institute NASU \& NSAU, \\ 
prospekt Akademika Glushkova 40 build. 4/1, \\ 03680 MSP Kyiv-187 Ukraine} 
\date{\today} 

\begin{document} 
 
\maketitle 
 
\begin{abstract}
On behalf of the International Women Day, the Sun gave a hot kiss to our mother 
Earth in a form of a full halo CME generated by the yesterday's double X-class 
flare. The resulting geomagnetic storm gives a good opportunity to compare the 
performance of space weather forecast models operating in near-real-time. We 
compare the forecasts of most major models and identify some common problems.
We also present the results of our own near-real-time forecast models.
\end{abstract} 

\section{Introduction}\label{s:Introduction}
The storm started yesterday with a side blow from a CME generated by the X1.1 
flare on March 5th after a long period of southward-directed IMF. This first G2 
storm scored the $K_P$ index of 6 and lasted over 12 hours. Around midnight on 
March 7th a huge X5.4 flare has spewed an Earth-directed CME, almost 
immediately causing an R3 blackout and an S3 radiation storm. It was soon 
followed by an X1.3 flare, but whether this one was followed by a CME is not 
clear. This is a very important issue, since all known superstorms were caused 
by several sequential CMEs. According to SOHO measurements at about 1000Z the 
solar wind velocity rapidly increased from 500 to 700~$\textrm{km\,s}^{-1}$ 
with a peak value of 800~$\textrm{km\,s}^{-1}$. Today at 1105Z the CME has hit 
the Earth causing an impressive 58nT sudden impulse and instantly pumping $K_P$ 
to 5. The $B_Z$ component was northward most of the time. For this reason, this 
storm is much weaker than it could be. If the IMF turns southward for a 
considerable length of time, it will become much more intense.

To add to forecasters' troubles, the ACE data was unavailable for most part of 
the day. Since the availability of near-real-time (NRT) solar wind data is critical 
for most space weather (SWx) forecasts, it is intriguing to see how the 
forecast models perform in such conditions. We will consider only the forecasts 
of the geomagnetical indices $D_{ST}$ and $K_P$, since these are the most 
widely used SWx parameters.

\section{Analysis of the models' performance}\label{s:Analysis}
We start with the $D_{ST}$ index, since it was historically the first proxy for SWx 
forecasting. The NRT online models include the empirical Temerin and Li model
(http://lasp.colorado.edu/space\_weather/dsttemerin/dsttemerin.html), and two 
models based on artificial neural networks: the Wintoft Dst model 
(http://rwc.lund.irf.se/rwc/dst/index.php) dating back to 2002 and the NICT 
model (http://www2.nict.go.jp/aeri/swe/swx/ace/nnw/) . All these models provide 
1 hour lead time, so they are nowcast rather then forecast models. Recently 
they were joined by a new empirical model by Podladchikova 
(http://www.spaceweather.ru/content/extended-geomagnetic-storm-forecast), which 
outputs not the $D_{ST}$  index itself, but an estimation of its peak value. For 
this reason, the lead time cannot be clearly defined for this model, but 
typically the storm is forecasted 2-3 hours before the commencement. Also, 
there is a model developed by the authors of this article, which will be 
integrated in the DLR's Space Weather Application Center Ionosphere 
(http://swaciweb.dlr.de/). It uses the regression modelling approach and 
provides 3 hours lead time.

Let us see how these models performed during this storm. Temerin and Li model 
overestimated the magnitude of the first storm by 50\%, and Wintoft Dst model 
totally missed the sudden impulse, but otherwise they provided a reasonable 
forecast. The NICT model overestimated the first storm's magnitude by 25\% and 
missed the sudden impulse. Podladchikova model overestimated the magnitude of 
the first storm by 25\%, and her peak value prediction approach does not give 
the information on the onset time. Our $D_{ST}$ model delayed with the storm 
onset prediction and missed the sudden impulse. Also it sometimes lags 1-2 
hours behind the Kyoto $D_{ST}$ (thus providing 1-2 hours lead time).

Next we switch to the $K_P$ index, which gained some popularity lately due to 
its apparent clarity. Unfortunately, the harsh reality is that the $K_P$ index 
is notoriously difficult to use both for research and operational purposes due 
to the lack of physical sense.

The state-of-the-art Wing model (http://www.swpc.noaa.gov/wingkp/) became 
inoperational since 0830Z yesterday and still shows no vital signs.
Wintoft Kp model (http://rwc.lund.irf.se/rwc/kp/index.php) stopped working 
about yesterday noon occasionally providing unrealistically low $K_P$ forecasts 
(less then 1 $K_P$ unit). Our $K_P$ model underestimated the storm intensity by 
almost 3 times but at least it worked through the whole event.

\section{Conclusion}\label{s:Conclusion}
All operational $K_P$ forecast models failed during the storm. The $D_{ST}$ 
forecast models appeared to be less affected by the lack of NRT data and 
provided better accuracy. Considering the more evident physical meaning of the 
$D_{ST}$ index, it seems reasonable to use it for global models, and to develop 
regional indices for specific areas, especially in high latitudes.

The problem of missing ACE data can be addressed by filling the gaps in data. 
Such software is currently being developed by the authors of this article. 
Hopefully, future spacecraft for solar wind monitoring will have some sort of 
radiation-resistant instruments providing at least some very rough NRT data. In 
situations like this one it is better to have very rough data on the solar wind 
velocity than to have no information at all.

\section{Acknowledgements} 
The research leading to these results has received funding from the European 
Commission's Seventh Framework Programme (FP7/2007-2013) under the grant 
agreement No. 263506 (AFFECTS project, www.affects-fp7.eu).

We acknowledge the Geoforschung Zentrum Potsdam for the near-real-time $K_P$ 
data, the Kyoto WDC for Geomagnetism, Kyoto for the near-real-time $D_{ST}$ 
data, and the Space Physics Data Facility at NASA Goddard Spaceflight Center 
for the OMNI data.
 
The OMNI data were obtained from the GSFC/SPDF/NSSDC FTP server at 
ftp://nssdcftp.gsfc.nasa.gov/pub/spacecraft\_data/omni

We thank the ACE MAG and SWEPAM instruments teams and the ACE Science Center 
for providing the ACE data

\begin{figure}[p]
\centerline{\includegraphics[width=\columnwidth,clip]{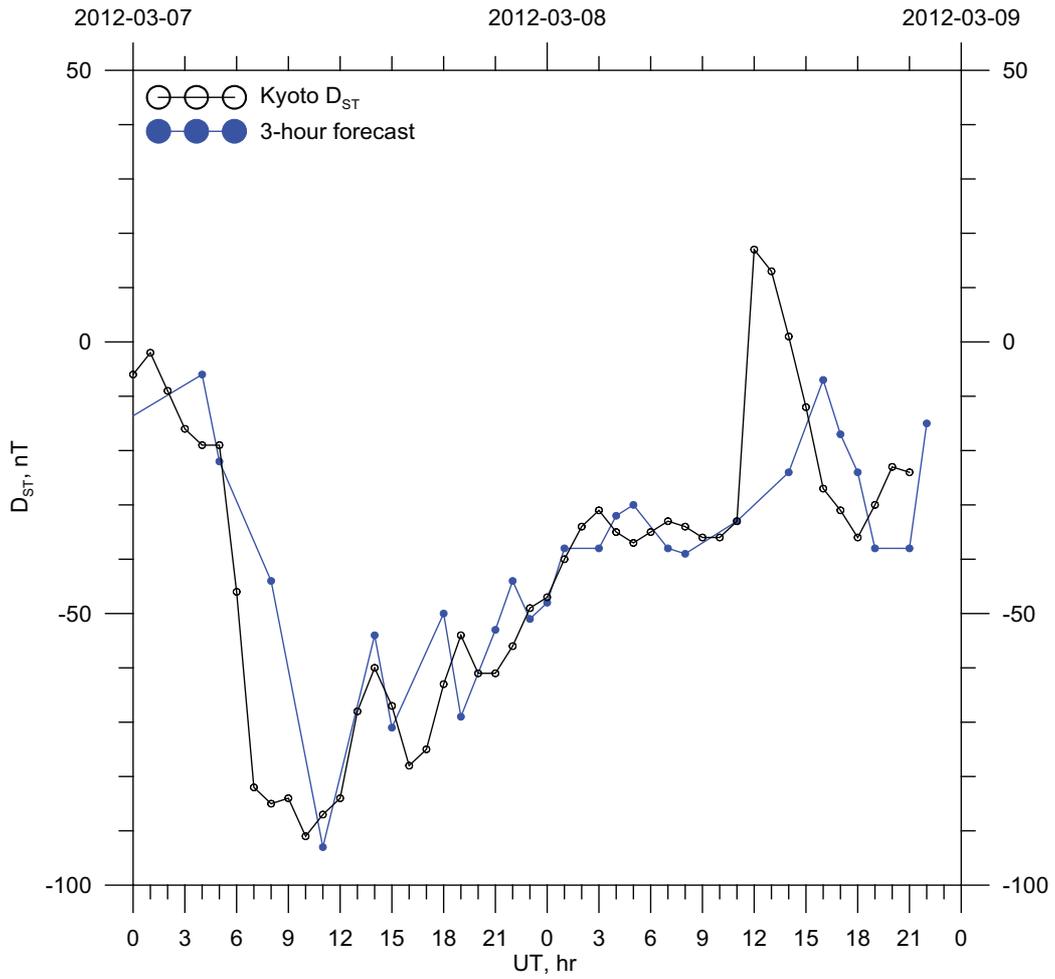}}
\caption{Our $D_{ST}$ model}
\label{fig:dst}
\end{figure}

\begin{figure}[p]
\centerline{\includegraphics[width=\columnwidth,clip]{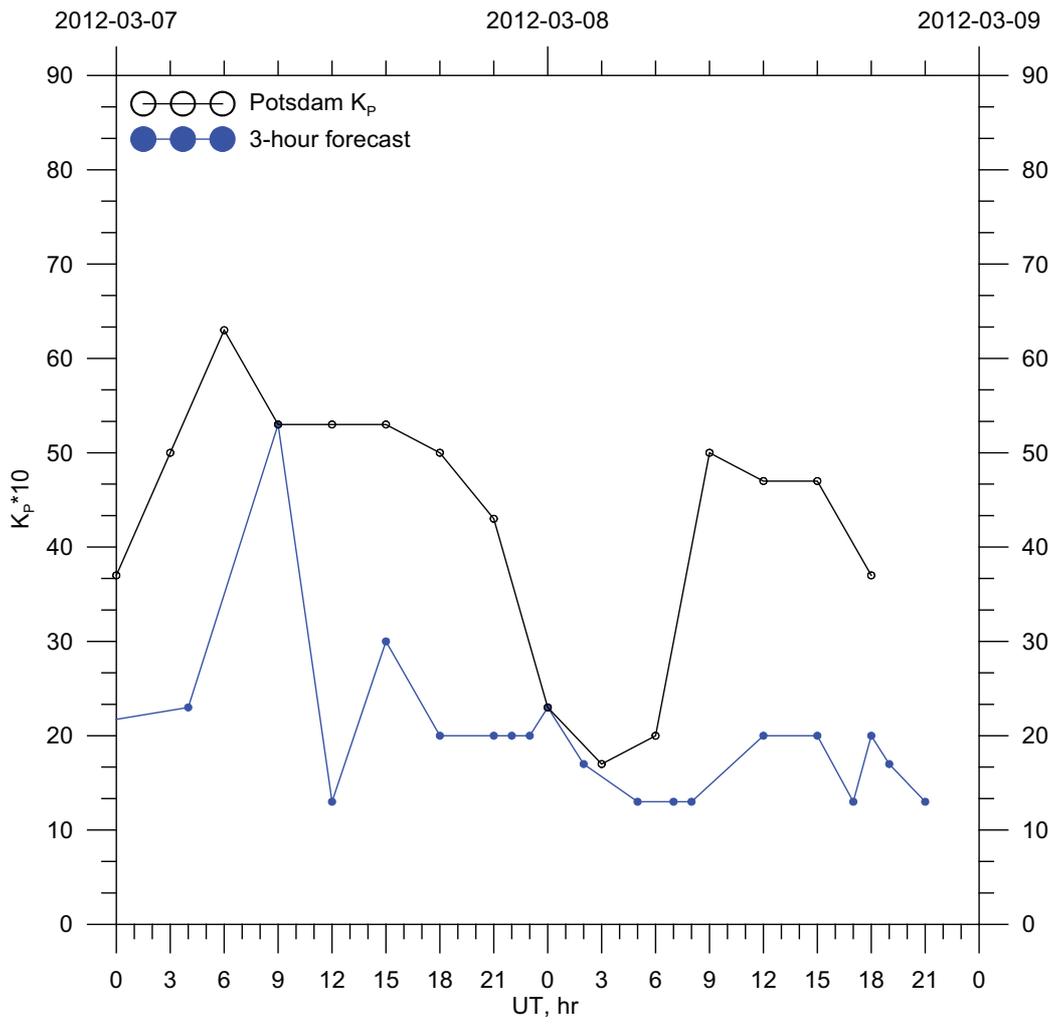}}
\caption{Our $K_P$ model}
\label{fig:kp}
\end{figure}

\end{document}